# Resonant diffraction and photoemission inconsistent with altermagnetism in epitaxial RuO₂ films

Benjamin Z. Gregory[1,2], Neha Wadehra[2], Shuyuan Zhang[1], Yi Wu[1], Samuel Poage[3], Jörg Strempfer[4], Asish K. Kundu[5], Anil Rajapitamahuni[5], Elio Vescovo[5], Anita Verma[2], Betül Pamuk[6], Jacob Ruf[1,7], Hari Nair[2], Nathaniel J. Schreiber[2], Kaveh Ahadi[3], Kyle M. Shen[1,8], Darrell G. Schlom[2,8,9], and Andrej Singer[2,*]

[1]*Laboratory of Atomic and Solid State Physics, Department of Physics, Cornell University, Ithaca, NY 14853, USA*
[2]*Department of Materials Science and Engineering, Cornell University, Ithaca, NY 14853, USA*
[3]*Department of Materials Science and Engineering, The Ohio State University, Columbus, OH 43210, USA*
[4]*Advanced Photon Source, Argonne National Laboratory, Lemont, IL 60439, USA*
[5]*National Synchrotron Light Source II, Brookhaven National Laboratory, Upton, NY 11973, USA*
[6]*Department of Physics and Astronomy, Williams College, Williamstown, MA 01267, USA*
[7]*Max-Planck Institute for Chemical Physics of Solids, Nöthnitzer Straße 40, 01187 Dresden, Germany*
[8]*Kavli Institute at Cornell for Nanoscale Science, Cornell University, Ithaca, NY 14853, USA*
[9]*Leibniz-Institut für Kristallzüchtung, Max-Born-Straße 2, 12489 Berlin, Germany*

**Abstract:**

Excitement about the magnetic and electronic properties of RuO₂ is growing, fueled by reports of antiferromagnetism, strain-induced superconductivity, and its recent classification as a member of a newly proposed magnetic class—altermagnets—with RuO₂ widely regarded as the paradigmatic example. Nevertheless, the magnetic ground state of RuO₂ remains contentious, as several recent experiments report no evidence of magnetic order. To address this discrepancy, we performed resonant elastic scattering measurements on a series of epitaxial RuO₂ thin films grown on the (100)-plane of TiO₂ substrates across a range of strain states. Leveraging full polarization control and azimuthal scans of the structurally forbidden 100 Bragg reflection, we systematically tested for signatures of colinear antiferromagnetic order. We found that the resonant elastic scattering signal in RuO₂ thin films likely originates from anisotropic charge scattering, not long-range antiferromagnetic order. Using angle-resolved photoemission spectroscopy we uncover a band structure without altermagnetic band splitting that is consistent with a nonmagnetic phase. Similarly, anisotropic magnetoresistance results show no evidence of magnetism. The combination of three independent measurements suggests the absence of altermagnetism in RuO₂.

**Main Text:**

Excitement is growing about the magnetic ground state of RuO₂, driven by reports of antiferromagnetism in bulk crystals [1,2] and strain-stabilized superconductivity in epitaxial films [3,4]. More recently, a new form of magnetism has been reported that shares similarities with antiferromagnetism but differs in how the spin sublattices are related by symmetry. In conventional antiferromagnets, one spin sublattice transforms into the other via translation or inversion, both of which preserve Kramers degeneracy, meaning that spin bands remain unsplit. In contrast, in the newly dubbed "altermagnets," spin-up and spin-down sites are related by rotation or rotoinversion

(but not inversion), which allows for spin splitting of the energy bands. RuO$_2$ has emerged as the paradigmatic example of this new magnetic order.

Theoretical predictions suggested that RuO$_2$ should exhibit an anomalous Hall effect [5] and further theoretical work predicted distinct spin current and spin-orbit torque responses [6,7]. These predictions have led to experimental efforts aimed at confirming the unique transport and symmetry-breaking signatures of altermagnetism in RuO$_2$. The anomalous Hall effect was observed in RuO$_2$ epitaxial films on different substrates [8]. Spin torque effects were further observed [9–11]. Direct spectroscopic evidence for time-reversal symmetry breaking was obtained via magnetic circular dichroism in angle-resolved photoemission spectra, indicating a spin split band structure, which is a hallmark of altermagnetic order [12].

While some experimental results revealed signatures of altermagnetism in the electronic structure and transport properties of RuO$_2$, the direct evidence of intrinsic antiferromagnec order rests on only two diffraction experiments [1,2], a subject of continuing debate. Spin-resolved photoemission spectroscopy studies of RuO$_2$ concluded the band structure is non-magnetic [13]. Recent work using muon spin resonance (μSR) found no evidence of any magnetic order in bulk RuO$_2$ [14], a finding supported by further neutron diffraction and μSR experiments [15,16] which attribute the alleged magnetic peak [2] to the Renninger effect, a mere artifact of the scattering geometry. Other authors have questioned the conclusiveness of the resonant X-ray diffraction results [1], proposing alternative candidates for the magnetic order and suggesting diffraction-based experiments to verify these predictions [17,18]. To resolve the disagreements over the nature of magnetism in RuO$_2$, we performed resonant elastic X-ray scattering measurements (REXS) on a series of thin films, leveraging full polarization analysis including linear a circular polarization. While the altermagnetism experiments have focused on thin films, and magnetic structure studies have primarily examined bulk crystals, we close the gap between these two regimes by analyzing the same ordering wavevector reported in bulk [1,2,16,19] in epitaxial thin films of RuO$_2$ over a range of strain states.

**REXS Measurements of Strained RuO$_2$ Films**
Bulk RuO$_2$ has a rutile crystal structure (space group #136 ($P4_2/mnm$), $a$ = 4.492 Å, $c$ = 3.106 Å) at room temperature. We report on a set of RuO$_2$ thin films synthesized via molecular-beam epitaxy (MBE) on the (100)-plane of isostructural TiO$_2$ single crystal substrates ($a$ = 4.594 Å, $c$ = 2.959 Å) with lattice mismatch and orientation chosen to maximize compressive strain along the $c$-axis [4]. Figure 1(a) shows the crystal structure of these epitaxial films, their nominal in-plane strains (-4.7% along [001], and +2.3% along [010]). The colinear antiferromagnetic order proposed in literature [1,2] features a Néel vector along [001], and magnetic moments of the ruthenium atoms alternating in sign between successive (100) planes. The thicknesses of the films in this study range from 3 nm to 64 nm. The films of intermediate thicknesses, which includes the 22 and 42 nm samples in this study, also exhibit strain-induced superconductivity below around 1 K [3,4]. We performed room-temperature X-ray diffraction measurements using 15.5 keV X-rays at the QM2 beamline at the Cornell High Energy Synchrotron Source (CHESS) to characterize the strain state of these films. Lattice constants for all samples included in this study are shown in Fig. 1(b) and show various epitaxial strain states which monotonically relax the $c$-axis compression with increasing thickness from 3 nm to 64 nm.

To clarify the present ambiguities of the available scattering results and investigate the effect of large c-axis strain on the potential magnetic structure of RuO2, we performed resonant elastic X-ray scattering experiments (Fig. 1(c)). Using the common linear polarization basis which denotes polarization normal to the scattering plane as $\sigma$ and parallel to the scattering plane as $\pi$, one can decompose the scattering contributions into four channels: $\sigma\sigma'$ ($\sigma$:incident, $\sigma'$: scattered), $\sigma\pi'$, $\pi\sigma'$, and $\pi\pi'$. The total scattering intensity can be formulated in terms of a generic scattering factor, $F$, from each of the four polarization channels using Stokes parameters $S_1, S_2, S_3$ to describe the polarization state of incident X-rays [20]

$$I \propto (|F_{\sigma\sigma'}|^2 + |F_{\sigma\pi'}|^2 + |F_{\pi\sigma'}|^2 + |F_{\pi\pi'}|^2) + S_1(|F_{\sigma\sigma'}|^2 + |F_{\sigma\pi'}|^2 - |F_{\pi\sigma'}|^2 - |F_{\pi\pi'}|^2) + 2S_2 \operatorname{Re}(F_{\sigma\sigma'}F_{\pi\sigma'}^* + F_{\sigma\pi'}F_{\pi\pi'}^*) + 2S_3 \operatorname{Im}(F_{\sigma\sigma'}F_{\pi\sigma'}^* + F_{\sigma\pi'}F_{\pi\pi'}^*). \quad (1)$$

Resonant elastic X-ray scattering couples to the element-specific resonance of core electronic transitions to probe long-range ordering of charge, orbital, and spin degrees of freedom [21,22]. Unlike non-resonant Thomson scattering, which primarily measures the density of core electrons and thus the ionic positions, REXS enhances sensitivity to anisotropies and multipole contributions near resonant energies. Polarization and azimuthal-angle dependence allow REXS to uncover symmetry-breaking order parameters, such as orbital or magnetic ordering. Specifically, REXS is sensitive to unit cell doubling due to electronic and magnetic order while the structure remains undoubled. As a result, structurally forbidden peaks—absent in Thomson scattering—become allowed at resonance. Detailed information on the electronic and magnetic order is accessible through variation of the incident and analysis of the scattered X-ray polarizations with respect to the crystal structure. Both polarizations can be varied by tuning the polarization of incident beam, the azimuthal rotation of the sample around the measured reciprocal lattice vector, and polarization analysis of the scattered X-rays. The polarization and azimuth dependence of structurally forbidden Bragg reflections have already provided refinement of the ordered phases in various ruthenate systems [1,19,23–27].

Assuming scattering from a collinear antiferromagnetic ground state as the origin of the REXS signal in RuO2, the functional form of azimuthal scans can be accounted for with the simplest model of resonant magnetic scattering by a magnetic ion in spherical symmetry. The modulus of the magnetic form factor of this scattering process is given by [28,29]

$$|f^{\text{mag}}|^2 \propto |\boldsymbol{\epsilon} \times \boldsymbol{\epsilon}' \cdot \hat{\boldsymbol{m}}|^2 \quad (2)$$

with $\boldsymbol{\epsilon}$ and $\boldsymbol{\epsilon}'$, the polarization states of the incident and scattered photons, respectively, and $\hat{\boldsymbol{m}}$ the magnetic moment unit vector. Even though the crystal field around the ruthenium atom has orthorhombic ($D_{2h}$) symmetry, the scattering tensor reduces to the same form as that in spherical symmetry when the Néel vector points along [001] [29]. In this model, the expected azimuthal dependence of intensities in each polarization channel follows

$$\left|F_{\sigma\sigma'}^{\text{mag}}(\psi)\right|^2 = 0,$$
$$\left|F_{\sigma\pi'}^{\text{mag}}(\psi)\right|^2 \propto \cos^2(\psi),$$
$$\left|F_{\pi\sigma'}^{\text{mag}}(\psi)\right|^2 \propto \cos^2(\psi),$$
$$\left|F_{\pi\pi'}^{\text{mag}}(\psi)\right|^2 \propto \sin^2(\psi),$$

where $\psi = 0°$ occurs when the RuO$_2$ [001] lies within the scattering plane (Fig. 1). The sinusoidal dependence has a straightforward interpretation: the vectors $\boldsymbol{\epsilon}_\sigma \times \boldsymbol{\epsilon}_{\pi'}$ and $\boldsymbol{\epsilon}_\pi \times \boldsymbol{\epsilon}_{\sigma'}$ both remain within the scattering plane and the dot product with $\hat{\boldsymbol{m}}$ results in the $\cos(\psi)$ term maximized at $\psi = 0$ for the $\sigma\pi'$ and $\pi\sigma'$ channels. In the $\pi\pi'$ channel, the vector $\boldsymbol{\epsilon}_\pi \times \boldsymbol{\epsilon}_{\pi'}$ is normal to the scattering plane, so the azimuthal intensity is maximized 90 degrees off from $\sigma\pi'$ and $\pi\sigma'$, resulting in the $\sin(\psi)$ dependence.

## REXS with Linear Polarization

We tuned the X-ray energy to the Ru $L_2$ edge (2.9685 keV) and measured azimuthal scans of the structurally forbidden yet resonantly allowed 100 Bragg peak at beam line 4-ID-D of the Advanced Photon Source at Argonne National Laboratory (see Supplementary Material for data on other X-ray energies). We first discuss the azimuthal dependence with linearly $\sigma$-polarized ($S_1 = 1, S_2 = 0, S_3 = 0$) incident X-rays and no polarization analyser, integrating the signal from both channels $\sigma\sigma'$ and $\sigma\pi'$. Such azimuthal scans for a range of epitaxial films are shown in Fig. 1(d). We make two observations from these data. The first is the uniformity of the azimuthal dependence across samples under radically different strain conditions. The second is the simple sinusoidal shape shared by all the samples (residuals in the Supplementary Material [30]), consistent with $I^{\text{mag}}(\psi) \propto |F^{\text{mag}}_{\sigma\sigma'}|^2 + |F^{\text{mag}}_{\sigma\pi'}|^2 \propto \cos^2(\psi)$ given by Eq. (3).

To resolve the individual contributions from the $\sigma\sigma'$ and the $\sigma\pi'$ channels, we inserted a GdMnO$_3$ polarization analyzer downstream of the scattered X-rays and obtained the 100 azimuthal dependence in the $\sigma\pi'$ channel for two samples, 9 nm and 42 nm, representing different strain relaxation conditions. These data are shown in Fig. 2(a) and are similarly well-described by Eq. (3). By rotating the analyzer crystal and the detector around the scattered beam by the angle $\eta$ in Fig. 1(c), we continuously tune the polarization of the measured X-rays, initially measuring pure $\sigma\pi'$ eventually measuring the pure $\sigma\sigma'$ scattering when $\eta = 0°$. The inset of Fig. 2(a) shows the analyzer angle dependence of the scattered intensity at the azimuthal maximum ($\psi = 0°$), obtained by rocking the analyzer through the GdMnO$_3$ 020 reflection with $2\theta = 45.41°$ and leakage from other polarization channel less than $2 \times 10^{-4}$. We observe decreasing intensity sweeping from $\sigma\pi'$ to $\sigma\sigma'$, which follows the $\sin^2(\eta)$ dependence expected if only $\pi'$ light is scattered from the sample. These data indicate negligible REXS scattering in the $\sigma\sigma'$ channel.

## REXS with Circular Polarization

With circular polarization, resonant X-rays simultaneously scatter in all four polarization channels. From Eq. (1) the sum of scattered intensities with right-handed ($I_+: S_1 = S_2 = 0, S_3 = 1$) and left-handed ($I_-: S_1 = S_2 = 0, S_3 = -1$) X-rays is proportional to the sum of intensities in each polarization channel

$$I_+ + I_- \propto |F_{\sigma\sigma'}|^2 + |F_{\sigma\pi'}|^2 + |F_{\pi\sigma'}|^2 + |F_{\pi\pi'}|^2.$$

Using a diamond phase plate in quarter wave plate condition upstream of the sample upstream of the sample (Fig. 1(c)), we measured the azimuthal dependence of the integral intensity of the 100 Bragg peak for three films (3 nm, 9 nm, and 42 nm) under both right- and left-handed circularly polarized X-rays (see Supplementary Material for the individual azimuthal scans) . The sums of these right- and left- handed azimuthal scans are shown in Fig. 2(b). Under the collinear antiferromagnetic model (Eq. (3)), the azimuthal dependence of the integral intensity of the resonant 100 Bragg peak is described by two sinusoidal contributions

$$I_+^{mag}(\psi) + I_-^{mag}(\psi) = a_1\cos^2(\psi) + a_2\sin^2(\psi)$$

with $a_1$ and $a_2$ as parameters depending on experimental geometry. Fits to the data with this expression are shown with dashed black curves in Fig. 2(b). For all measured samples, $a_2$ is found to be zero. The absence of the $\sin^2(\psi)$ signal contradicts the predicted $\pi\pi'$ scattering under the collinear antiferromagnetic model (Eq. (3)).

Recent theoretical work has expressed skepticism at the conclusion [1] that collinear antiferromagnetism is present in RuO$_2$, highlighting the difficulty of conclusively identifying the magnetic origin of an azimuthal signal without access to all four possible linear polarization scattering channels [17,18]. Those authors forecast that charge-magnetic interference will occur in resonant scattering from RuO2, which can be exposed using circular polarization analysis. They propose possible antiferromagnetic structure motifs, which are not ruled out by our preceding analysis, but would produce a "chiral signature" in their diffraction, an unambiguous indication of long-range antiferromagnetic order. The chiral signature is defined in Ref. [17] as,

$$\gamma = \text{Im}(F_{\sigma\sigma'}F_{\pi\sigma'}^* + F_{\sigma\pi'}F_{\pi\pi'}^*) = I_+ - I_-,$$

and can reveal interferences between channels. Figure 2(c) shows the chiral signature, $\gamma$, explicitly for the same three samples as the difference between the left- and right-handed data points plotted on the same scale. We observe only statistical fluctuations near zero for all three films, with larger variance for the ultrathin 3 nm due to low signal, consistent with $|F_{\sigma\sigma'}|^2 = |F_{\pi\pi'}|^2 = 0$ as discussed above. This "null" chiral signature indicates the absence of charge-magnetic interference and, to our experimental resolution, rules out the three magnetic motifs proposed in Ref. [17].

The combination of scattering in the $\sigma\pi'$ or $\pi\sigma'$ scattering and the absence of scattering in the $\sigma\sigma'$ channel is regarded as a hallmark of antiferromagnetic order, a claim often supported by neutron diffraction [24,27,31–34]. Using polarization-controlled azimuthal scans of the structurally forbidden 100 Bragg reflection, we found that our linear polarization data (Figs. 1 and 2(a)) directly reveal a simple $\cos^2(\psi)$ dependence in the $\sigma\pi'$ channel and show no $\sigma\sigma'$ scattering. Yet data collected with circular polarization (Figs. 2(b,c)) show no evidence of the expected antiferromagnetic $\pi\pi'$ scattering. Therefore, the currently accepted model of RuO$_2$ antiferromagnetism, which arose to account for the observation of the 100 peak with neutron diffraction [2] and REXS [1] in bulk RuO$_2$, is inconsistent with our azimuthal scans. These data cannot rule out definitively the presence of a weak antiferromagnetic signal in our films, which may be suppressed due to miniscule moments or small domains, but they do show that the collinear antiferromagnetism cannot be the primary source of the resonant 100 peak.

**Anisotropic Tensor of Susceptibility**
A better explanation of our REXS data is anisotropic tensor of susceptibility (ATS) scattering, also known as Templeton-Templeton scattering [35,36], which arises from local anisotropy of the X-ray susceptibility tensor around the resonant ion during the REXS process. On resonance this anisotropy can produce observable diffraction intensity on structurally forbidden reflections with polarization dependence similar to resonant magnetic scattering [37]. TiO$_2$ provided an early example of a crystal where the octahedral oxygen coordination of titanium in the rutile structure was predicted to recover "forbidden" intensity of the 100 Bragg reflection on resonance [38]. Later experiments realized these predictions at the Ti K edge, observing a simple sinusoidal azimuthal dependence of the 100 peak with incident sigma X-rays [39,40]. The resonant intensity of the 100 peak in TiO$_2$ arises from the local $D_{2h}$ symmetry around the titanium atom, which leads to zero intensity in the $\sigma\sigma'$ channel and a simple $\cos^2(\psi)$ dependence in $\sigma\pi'$ (see Eq. (40) in Ref. [39]).

Even with an orthorhombic distortion of the RuO$_2$ unit cell, the oxygen coordination around ruthenium retains $D_{2h}$ symmetry and thus can be expected to produce similar ATS scattering at the resonant 100 peak as in TiO$_2$. Assuming an orthorhombic ($D_{2h}$) coordination environment around the ruthenium ion, we expect the following azimuthal dependences of ATS scattering on the 100 peak [30]:

$$\begin{aligned}|F^{ATS}_{\sigma\sigma'}(\psi)|^2 &= 0, \\ |F^{ATS}_{\sigma\pi'}(\psi)|^2 &\propto \cos^2(\psi), \\ |F^{ATS}_{\pi\sigma'}(\psi)|^2 &\propto \cos^2(\psi), \\ |F^{ATS}_{\pi\pi'}(\psi)|^2 &= 0.\end{aligned} \quad (4)$$

The ATS model fits our azimuthal data better, which all are adequately described by $\cos^2(\psi)$ in the $\sigma\pi'$ and $\pi\sigma'$ channels, and no contributions from other two channels. This model also reconciles the presence of very bright resonant Bragg peaks, even in 3 nm films, with reports of vanishingly small magnetic moments [14,15].

**ARPES Data**

To further examine the possible altermagnetism in epitaxial RuO$_2$ thin films on TiO$_2$, we performed angle-resolved photoemission spectroscopy (ARPES) measurements on RuO$_2$ films grown on both TiO$_2$ (100) and TiO$_2$ (110) substrates (Fig. 3). In altermagnetic RuO$_2$, the electronic band structures are predicted to exhibit a significant spin splitting along the $\Gamma$-$M$ and $A$-$Z$ directions [12,13,41–44]. Our ARPES measurements focused on momentum regions near these directions and were compared with density functional theory (DFT) calculations. DFT calculations were performed with the PBEsol+U level of the theory with a Hubbard $U = 2$ eV, including spin-orbit-coupling but excluding the altermagnetic order [3,4]. Figure 3(a) illustrates the schematic of the Brillouin zone of RuO$_2$ (100) films. The band structures along cut 1 (Fig. 3(b)), crossing with the $\Gamma$-$M$ direction, and along cut 2 (Fig. 3(c)), crossing with the $A$-$Z$ direction, match well with the DFT calculations without the altermagnetic order. Since the splitting could not be significant along the (100) surface, we also performed ARPES measurements on RuO$_2$ (110) films, presented in Fig. 3(d) to (f). Clear dispersing bands are observed along both $\Gamma$-$M$ (Fig. 3(e)) and $A$-$Z$ (Fig. 3(f)) directions. The DFT calculations reproduce the band dispersion well without considering altermagnetism, indicating that altermagnetic order is not essential in describing the observed band structures, consistent with [13].

**Anisotropic Magnetoresistance**

Anisotropic magnetoresistance (AMR) can be harnessed as a probe into magnetic structure and its evolution with applied magnetic field in itinerant antiferromagnets [45–48]. Figure 4 shows the AMR(%)= $(R_{xx}–R_{min})/R_{min} \times 100$, where the current is kept along a crystallographic direction as the angle between the current and the in-plane magnetic field changes. Here, we measured magnetoresistance in two $RuO_2$ films, 17 and 10 nm, grown on $TiO_2$ (100) and $TiO_2$ (101), respectively. The measurement was repeated for each sample along two perpendicular current directions. We note that the film grown on $TiO_2$ (101) shows larger magnetoresistance and there is a small dependence on the current direction. We attribute these to anisotropy of the epitaxial strain in these films [3,4].

AMR response in an itinerant antiferromagnet contains both crystalline and non-crystalline contributions [49]. The crystalline component depends on the angle between the Néel vector and a specific crystallographic direction. Here, the two-fold symmetry of the AMR matches the in-plane symmetry of the underlying lattice. The 90° degree rotation of the current direction, however, does not cause a corresponding rotation in the AMR response for films grown on $TiO_2$ (100) and $TiO_2$ (101) which suggests that the observed magnetoresistance is not crystalline. The non-crystalline AMR, to first order, depends on how the scattering matrix elements of the charge carriers change with the angle between the spin axis and the momentum [45,49]. Here, again the change in applied magnetic field, film growth orientation, and current direction do not show a change in the AMR response. We conclude that the magnetoresistance results do not exhibit signs of magnetic ordering in $RuO_2$ films, and it is likely due to an ordinary, Hall-type magnetoresistance contribution.

**Discussion and Conclusions**
Our resonant elastic X-ray scattering study of epitaxially-strained $RuO_2$ thin films reveals a consistent azimuthal dependence of the structurally forbidden 100 reflection, well-described by a simple $\cos^2(\psi)$ form in the $\sigma\pi'$ channel, and with a vanishing intensity in the $\sigma\sigma'$ channel. Although this linear polarization dependence is characteristic of resonant magnetic scattering, our circular polarization analysis finds the expected magnetic $\pi\pi'$ contribution is absent, ruling out collinear antiferromagnetism as the origin of the resonant 100 Bragg peak. Instead, our data are better explained by anisotropic tensor of susceptibility (ATS) scattering, arising from the $D_{2h}$ point symmetry of the ruthenium ions. The ATS model alone fully accounts for the observed polarization and azimuthal dependence of the resonant reflection. With ARPES, we find no evidence of band splitting in $RuO_2$ on (110)-$TiO_2$ and (100)-$TiO_2$ substrates, indicating the absence of altermagnetism in these films. Likewise, magnetoresistance measurements show no dependence of the AMR signal on applied magnetic field or current direction, consistent with a nonmagnetic metal. Consistent with recent μSR, neutron scattering, and spectroscopic results suggesting the absence of magnetism in $RuO_2$, our findings suggest that the resonant 100 peak originates from anisotropic charge scattering, not long-range magnetic order. Our diffraction, spectroscopic, and magnetoresistance results add to the evidence from other techniques [13–16] that magnetism is absent in $RuO_2$. The remaining experimental discrepancies are in the neutron scattering studies.

Berlijn *et al.* originally observed the 100 reflection and reported antiferromagnetism [2]. More recent work has argued that the observed 100 peak does not arise from magnetic order or a structural distortion, but rather is the result of multiple scattering known as the Renninger effect [15,16], which is observable in azimuthal scans as a sharp spike in intensity when two reflections briefly satisfy the Bragg condition simultaneously. Our data, however, show a smooth, sinusoidal azimuthal dependence, unlike a Renninger scan, and the presence of the 100 peak for bulk samples [1] and films in multiple strain states and with multiple substrate orientations (100), (110) and (101) [19], is very unlikely to arise from multiple scattering, which is highly sensitive to strain. We conclude that the resonant 100 X-ray peak is not due to the Renninger effect.

The original neutron study [2] reported a very small magnetic moment and recent muon spin resonance studies have pushed the upper bound of the magnetic moments down so low that they reasonably conclude the moments in bulk $RuO_2$ are zero [14], with a small increase of magnetic signal observed in thin films grown on $TiO_2$ (110), possibly arising from the epitaxial strain [15]. The failure of multiple scattering to account for the resonant 100 reflection and the observation of enhanced magnetic µSR signal in thin films leave open the possibility of weak antiferromagnetism. Reports of spin hall conductivity measurements find a Néel vector in $RuO_2$ films along the *c*-axis [9]. Evidence of time-reversal symmetry breaking was reported from photoemission experiments [12], pointing to magnetism in thin films. It could be that a small magnetic signal exists, and adds to the ATS component in our data, but if this is the case, the magnetic contribution never rises above the noise. It is likely there is only one source of resonant X-ray signal and that is anisotropy.

**Acknowledgments**
We acknowledge helpful conversations with C. A. Occhialini and J. Pelliciari. The work was primarily supported by U.S. Department of Energy, Office of Science, Office of Basic Energy Sciences, under Contract No. DE-SC0019414 (X-ray experiments and interpretation B.Z.G., J.S., A.V., A.S.; thin film synthesis: N.W., H.N., N.J.S). This research used resources of the Advanced Photon Source, a U.S. Department of Energy (DOE) Office of Science User Facility operated for the DOE Office of Science by Argonne National Laboratory under Contract No. DE-AC02-06CH11357. Part of ARPES measurements were performed at the ESM beamline of the National Synchrotron Light Source II, a U.S. Department of Energy (DOE) Office of Science User Facility operated for the DOE Office of Science by Brookhaven National Laboratory under Contract No. DE-SC0012704. This work was also funded in part by the Gordon and Betty Moore Foundation's EPiQS Initiative, Grant GBMF9073 to Cornell University to support the work of D.G.S, and NSF DMR-22104427 and AFOSR FA9550-21-1-0168 (characterization and model development, J.R., K.M.S.).

**Data availability**

The Resonant Elastic X-ray Scattering Data presented in this work are available for download at [link to be included at the date of the publication].

# References

[1] Z. H. Zhu, J. Strempfer, R. R. Rao, C. A. Occhialini, J. Pelliciari, Y. Choi, T. Kawaguchi, H. You, J. F. Mitchell, Y. Shao-Horn, and R. Comin, Phys. Rev. Lett. **122**, 17202 (2019).

[2] T. Berlijn, P. C. Snijders, O. Delaire, H. D. Zhou, T. A. Maier, H. B. Cao, S. X. Chi, M. Matsuda, Y. Wang, M. R. Koehler, P. R. C. Kent, and H. H. Weitering, Phys. Rev. Lett. **118**, 2 (2017).

[3] J. P. Ruf, H. Paik, N. J. Schreiber, H. P. Nair, L. Miao, J. K. Kawasaki, J. N. Nelson, B. D. Faeth, Y. Lee, B. H. Goodge, B. Pamuk, C. J. Fennie, L. F. Kourkoutis, D. G. Schlom, and K. M. Shen, Nat. Commun. **12**, 59 (2021).

[4] N. Wadehra, B. Z. Gregory, S. Zhang, N. Schnitzer, Y. Iguchi, Y. E. Li, B. Pamuk, D. A. Muller, A. Singer, K. M. Shen, and D. G. Schlom, Commun. Mater. **6**, 135 (2025).

[5] L. Šmejkal, R. González-Hernández, T. Jungwirth, and J. Sinova, Sci. Adv. **6**, (2020).

[6] R. González-Hernández, L. Šmejkal, K. Výborný, Y. Yahagi, J. Sinova, T. Jungwirth, and J. Železný, Phys. Rev. Lett. **126**, 1 (2021).

[7] L. Šmejkal, J. Sinova, and T. Jungwirth, Phys. Rev. X **12**, 031042 (2022).

[8] Z. Feng, X. Zhou, L. Šmejkal, L. Wu, Z. Zhu, H. Guo, R. González-Hernández, X. Wang, H. Yan, P. Qin, X. Zhang, H. Wu, H. Chen, Z. Meng, L. Liu, Z. Xia, J. Sinova, T. Jungwirth, and Z. Liu, Nat. Electron. **5**, 735 (2022).

[9] A. Bose, N. J. Schreiber, R. Jain, D.-F. Shao, H. P. Nair, J. Sun, X. S. Zhang, D. A. Muller, E. Y. Tsymbal, D. G. Schlom, and D. C. Ralph, Nat. Electron. **5**, 267 (2022).

[10] H. Bai, L. Han, X. Y. Feng, Y. J. Zhou, R. X. Su, Q. Wang, L. Y. Liao, W. X. Zhu, X. Z. Chen, F. Pan, X. L. Fan, and C. Song, Phys. Rev. Lett. **128**, 197202 (2022).

[11] S. Karube, T. Tanaka, D. Sugawara, N. Kadoguchi, M. Kohda, and J. Nitta, Phys. Rev. Lett. **129**, 137201 (2022).

[12] O. Fedchenko, J. Minár, A. Akashdeep, S. W. D'Souza, D. Vasilyev, O. Tkach, L. Odenbreit, Q. Nguyen, D. Kutnyakhov, N. Wind, L. Wenthaus, M. Scholz, K. Rossnagel, M. Hoesch, M. Aeschlimann, B. Stadtmüller, M. Kläui, G. Schönhense, T. Jungwirth, A. B. Hellenes, G. Jakob, L. Šmejkal, J. Sinova, and H. J. Elmers, Sci. Adv. **10**, 1 (2024).

[13] J. Liu, J. Zhan, T. Li, J. Liu, S. Cheng, Y. Shi, L. Deng, M. Zhang, C. Li, J. Ding, Q. Jiang, M. Ye, Z. Liu, Z. Jiang, S. Wang, Q. Li, Y. Xie, Y. Wang, S. Qiao, J. Wen, Y. Sun, and D. Shen, Phys. Rev. Lett. **133**, 176401 (2024).

[14] M. Hiraishi, H. Okabe, A. Koda, R. Kadono, T. Muroi, D. Hirai, and Z. Hiroi, Phys. Rev. Lett. **132**, 166702 (2024).

[15] P. Keßler, L. Garcia-Gassull, A. Suter, T. Prokscha, Z. Salman, D. Khalyavin, P. Manuel, F. Orlandi, I. I. Mazin, R. Valentí, and S. Moser, Npj Spintron. **2**, 50 (2024).

[16] L. Kiefer, F. Wirth, A. Bertin, P. Becker, L. Bohatý, K. Schmalzl, A. Stunault, J. A. Rodríguez-Velamazán, O. Fabelo, and M. Braden, Arxiv:2410.05850 (2024).

[17] S. W. Lovesey, D. D. Khalyavin, and G. van der Laan, Phys. Rev. B **105**, 1 (2022).

[18] S. W. Lovesey, D. D. Khalyavin, and G. van der Laan, Phys. Rev. B **108**, 2 (2023).

[19] B. Z. Gregory, J. Strempfer, D. Weinstock, J. P. Ruf, Y. Sun, H. Nair, N. J. Schreiber, D. G. Schlom, K. M. Shen, and A. Singer, Phys. Rev. B **106**, 1 (2022).

[20] Y. Joly, Y. Tanaka, D. Cabaret, and S. P. Collins, Phys. Rev. B **89**, 1 (2014).

[21] J. P. Hill and D. F. McMorrow, Acta Crystallogr. Sect. A **52**, 236 (1996).
[22] J. Fink, E. Schierle, E. Weschke, and J. Geck, Reports Prog. Phys. **76**, (2013).
[23] B. Bohnenbuck, I. Zegkinoglou, J. Strempfer, C. S. Nelson, H.-H. Wu, C. Schü\ssler-Langeheine, M. Reehuis, E. Schierle, P. Leininger, T. Herrmannsdörfer, J. C. Lang, G. Srajer, C. T. Lin, and B. Keimer, Phys. Rev. Lett. **102**, 37205 (2009).
[24] B. Bohnenbuck, I. Zegkinoglou, J. Strempfer, C. Schü\ssler-Langeheine, C. S. Nelson, P. Leininger, H.-H. Wu, E. Schierle, J. C. Lang, G. Srajer, S. I. Ikeda, Y. Yoshida, K. Iwata, S. Katano, N. Kikugawa, and B. Keimer, Phys. Rev. B **77**, 224412 (2008).
[25] M. A. Hossain, I. Zegkinoglou, Y.-D. Chuang, J. Geck, B. Bohnenbuck, A. G. C. Gonzalez, H.-H. Wu, C. Schüßler-Langeheine, D. G. Hawthorn, J. D. Denlinger, R. Mathieu, Y. Tokura, S. Satow, H. Takagi, Y. Yoshida, Z. Hussain, B. Keimer, G. A. Sawatzky, and A. Damascelli, Sci. Rep. **3**, 2299 (2013).
[26] D. G. Porter, V. Granata, F. Forte, S. Di Matteo, M. Cuoco, R. Fittipaldi, A. Vecchione, and A. Bombardi, Phys. Rev. B **98**, 125142 (2018).
[27] I. Zegkinoglou, J. Strempfer, C. S. Nelson, J. P. Hill, J. Chakhalian, C. Bernhard, J. C. Lang, G. Srajer, H. Fukazawa, S. Nakatsuji, Y. Maeno, and B. Keimer, Phys. Rev. Lett. **95**, 136401 (2005).
[28] J. P. Hannon, G. T. Trammell, M. Blume, and D. Gibbs, Phys. Rev. Lett. **61**, 1245 (1988).
[29] M. W. Haverkort, N. Hollmann, I. P. Krug, and A. Tanaka, Phys. Rev. B **82**, 94403 (2010).
[30] Supplementary Material (n.d.).
[31] K. Gautam, S. S. Majid, S. Francoual, A. Ahad, K. Dey, M. C. Rahn, R. Sankar, F. C. Chou, and D. K. Shukla, Phys. Rev. B **101**, 224430 (2020).
[32] R. Caciuffo and C. Fisica, **65**, 1 (2002).
[33] V. Scagnoli, U. Staub, A. M. Mulders, M. Janousch, G. I. Meijer, G. Hammerl, J. M. Tonnerre, and N. Stojic, Phys. Rev. B - Condens. Matter Mater. Phys. **73**, 1 (2006).
[34] L. Paolasini, C. Vettier, F. De Bergevin, F. Yakhou, D. Mannix, A. Stunault, W. Neubeck, M. Altarelli, M. Fabrizio, P. A. Metcalf, and J. M. Honig, 4719 (1999).
[35] D. H. Templeton and L. K. Templeton, Acta Crystallogr. Sect. A **41**, 133 (1985).
[36] B. Y. D. H. Templeton and L. K. Templeton, **3**, 62 (1982).
[37] V. E. Dmitrienko, K. Ishida, A. Kirfel, and E. N. Ovchinnikova, Acta Crystallogr. Sect. A Found. Crystallogr. **61**, 481 (2005).
[38] V. E. Dmitrienko, Acta Crystallogr. Sect. A **39**, 29 (1983).
[39] A. Kirfel, A. Petcov, and K. Eichhorn, Acta Crystallogr. Sect. A **47**, 180 (1991).
[40] H. Sawai, J. Kokubun, and K. Ishida, **47**, 3114 (2003).
[41] R. M. Sattigeri, G. Cuono, and C. Autieri, Nanoscale **15**, 16998 (2023).
[42] A. Ptok, Arxiv:2309.02421 (2023).
[43] L. Šmejkal, A. B. Hellenes, R. González-Hernández, J. Sinova, and T. Jungwirth, Phys. Rev. X **12**, 11028 (2022).
[44] Z. Lin, D. Chen, W. Lu, X. Liang, S. Feng, K. Yamagami, J. Osiecki, M. Leandersson, B. Thiagarajan, J. Liu, C. Felser, and J. Ma, Arxiv 2402.04995 (2024).
[45] X. Marti, I. Fina, C. Frontera, J. Liu, P. Wadley, Q. He, R. J. Paull, J. D. Clarkson, J. Kudrnovský, I. Turek, J. Kuneš, D. Yi, J.-H. Chu, C. T. Nelson, L. You, E. Arenholz, S. Salahuddin, J. Fontcuberta, T. Jungwirth, and R. Ramesh, Nat. Mater. **13**, 367 (2014).
[46] K. Ahadi, X. Lu, S. Salmani-Rezaie, P. B. Marshall, J. M. Rondinelli, and S. Stemmer, Phys. Rev. B **99**, 041106 (2019).

[47] C. Wang, H. Seinige, G. Cao, J.-S. Zhou, J. B. Goodenough, and M. Tsoi, Phys. Rev. X **4**, 041034 (2014).

[48] C. Lu, B. Gao, H. Wang, W. Wang, S. Yuan, S. Dong, and J. Liu, Adv. Funct. Mater. **28**, 1 (2018).

[49] D. Kriegner, K. Výborný, K. Olejník, H. Reichlová, V. Novák, X. Marti, J. Gazquez, V. Saidl, P. Němec, V. V. Volobuev, G. Springholz, V. Holý, and T. Jungwirth, Nat. Commun. **7**, 11623 (2016).

# Figures

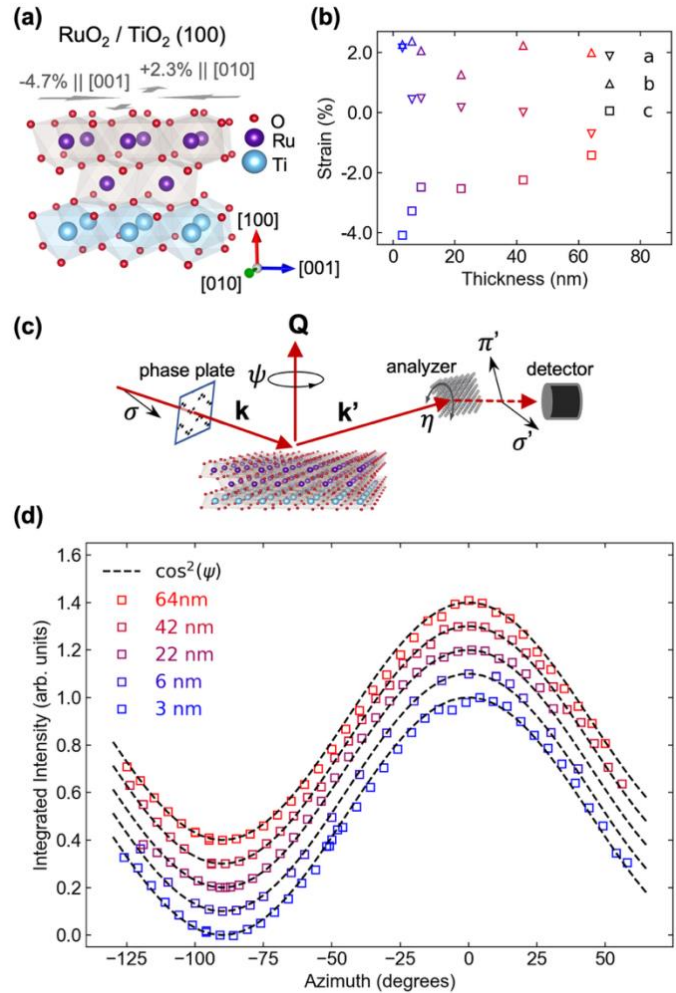

**Figure 1: Resonant elastic X-ray scattering from epitaxial RuO₂ films.** (a) Crystal structure of $RuO_2$ with nominal in-plane strains (gray arrows) imposed by the $TiO_2$(100) substrate. (b) Thickness-dependent lattice constants, showing the transition from commensurately strained to substantially relaxed. (c) Schematic of the resonant diffraction geometry: incident and scattered wavevectors **k** and **k'** defining the scattering plane, scattering vector **Q**, and azimuthal angle $\psi$ measured around **Q**, with $\psi = 0°$ when the $RuO_2$ c-axis lies within the scattering plane. The nominally linear polarization normal to the scattering plane ($\sigma$) can be changed to circular via a phase plate. A $GdMnO_3$ crystal can be used as polarization analyzer tuned to its 020 Bragg condition. Rotating the analyzer crystal and the detector around the scattered wavevector **k'** by the angle $\eta'$ controls how much radiation of each polarization state, $\sigma'$ or $\pi'$, arrives at the detector. (d) Azimuthal dependence of the integrated total intensity of the resonant 100 peak at 2.9685 keV with no phase plate or crystal analyzer. The dashed black curves show $\cos^2(\psi)$. All intensities are normalized to and vertically offset for clarity.

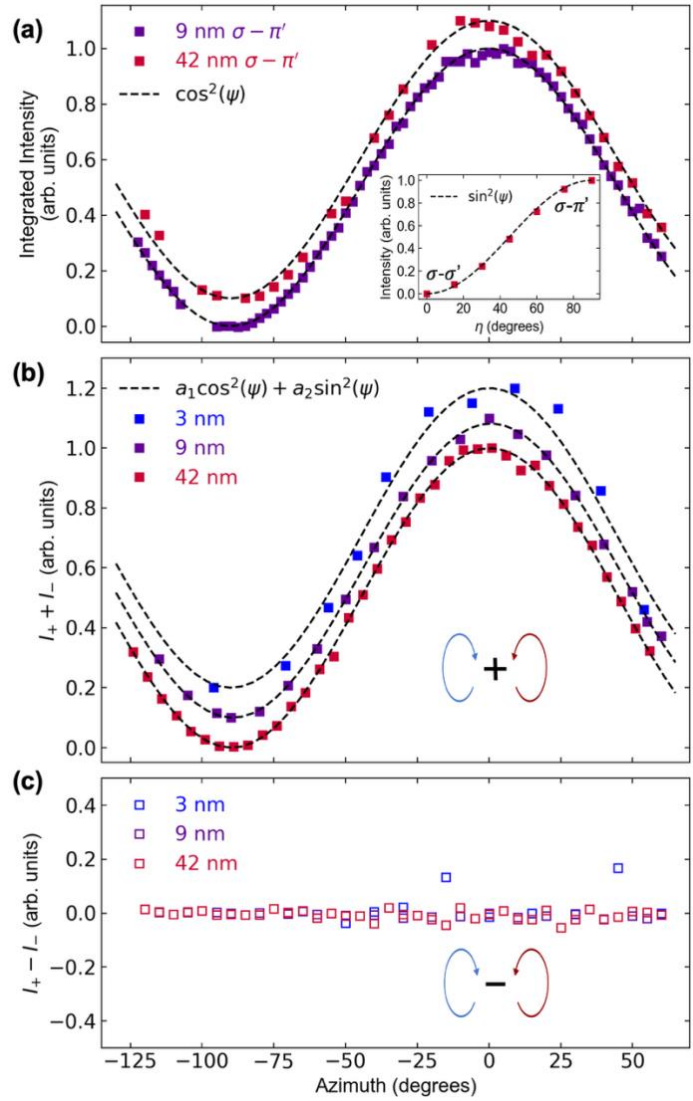

**Figure 2: Azimuthal scans for different incident polarizations of the resonant 100 Bragg reflection**. (a) Integrated intensity as a function of the azimuth in the $\sigma\pi'$ channel of a commensurately-strained film (9 nm) and a partially-relaxed film (42 nm). The dashed black curve shows $\cos^2(\psi)$. Inset: Integrated intensity of the analyzer 020 rocking curve as a function of analyzer angle ($\eta$ in Fig. 1(c)) while maintaining the 100 Bragg condition of the 42 nm sample with $\psi = 0°$. Intensity reduces from the $\sigma\pi'$ channel ($\eta = 90°$) to the $\sigma\sigma'$ channel ($\eta = 0°$). The dashed black curve is the theoretical response of purely $\sigma\pi'$ polarized radiation scattered from the sample. (b) Sum of integral intensities from right- and left-handed incident X-rays with no polarization analyzer. The dashed line is a fit of $a_1\cos^2(\psi) + a_2\sin^2(\psi)$. For the 3 nm, 9 nm, and 42 nm samples $a_1$ is 1.00(2), 1.000(2), 1.000(4), and $a_2$ is 0.000(6), 0.0001(6), 0.000(4), respectively. Data and fits are offset vertically for clarity. (c) The chiral signature, $\gamma$, obtained from the difference of left- and right-handed peak integral intensities shown in (b). In (a,b) the intensities are normalized to unity; data and theory curves vertically offset for clarity.

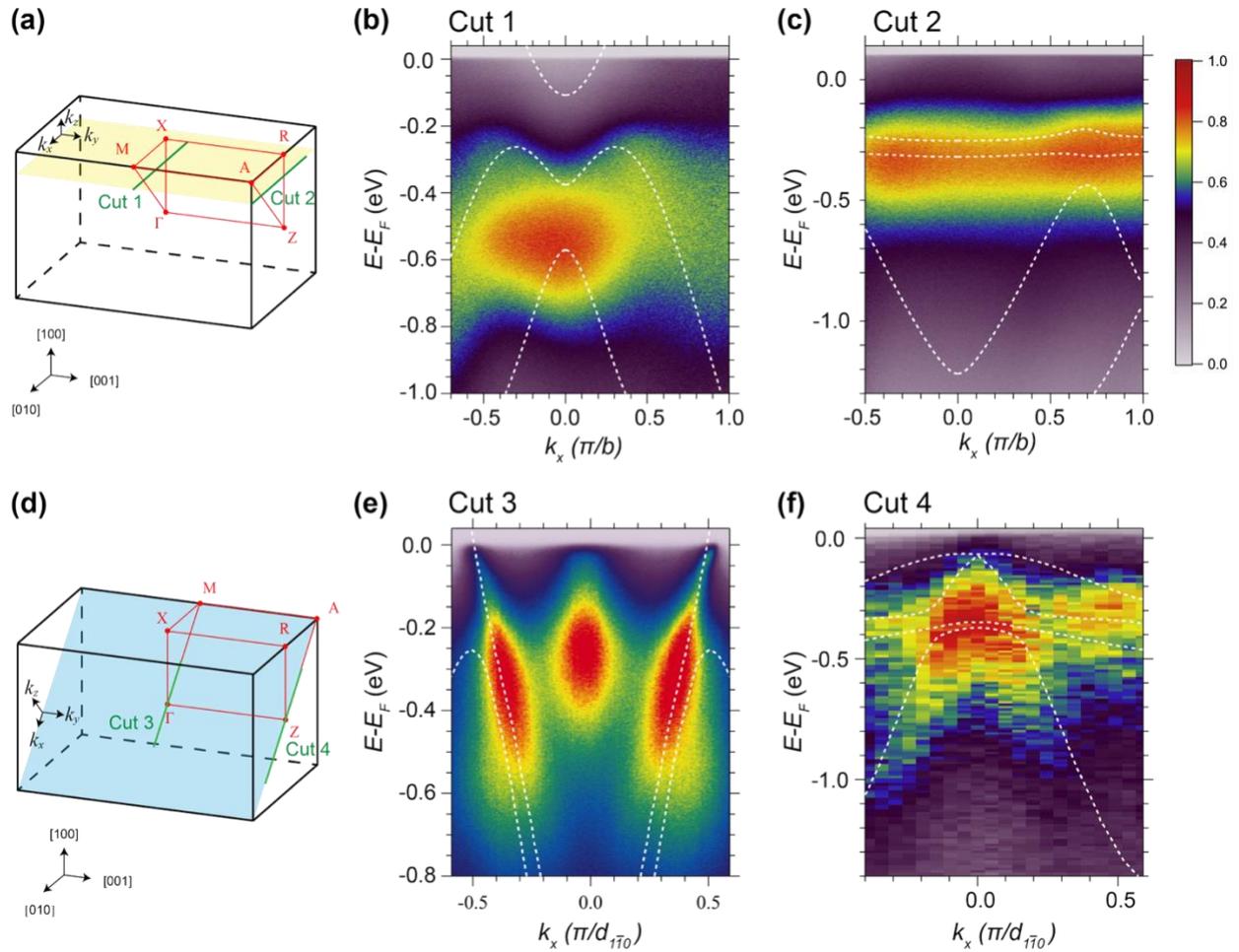

**Figure 3: Band structures of RuO₂ thin films measured via ARPES.** (a) Schematic of the Brillouin zone of RuO$_2$ (100) films. The yellow shading corresponds to the measured momentum plane at $k_z = 0.7$ $\pi/a$ using He-Iα photons (21.2eV). (b, c) Energy versus momentum spectra along (b) cut 1 at $k_y = 0$ and (c) cut 2 at $k_y = \pi/c$, as marked in (a). (d) Schematic of the Brillouin zone of RuO$_2$ (110) films. The blue shading corresponds to the measured momentum plane at $k_z = 0$. (e, f) Energy versus momentum spectra along (e) cut 3 at $k_y = 0$ using He-Iα photons (21.2eV) and (f) cut 4 at $k_y = \pi/c$ using 115 eV photons, as marked in (d). White dashed lines overlaying the ARPES spectra represent DFT calculations without the altermagnetic order.

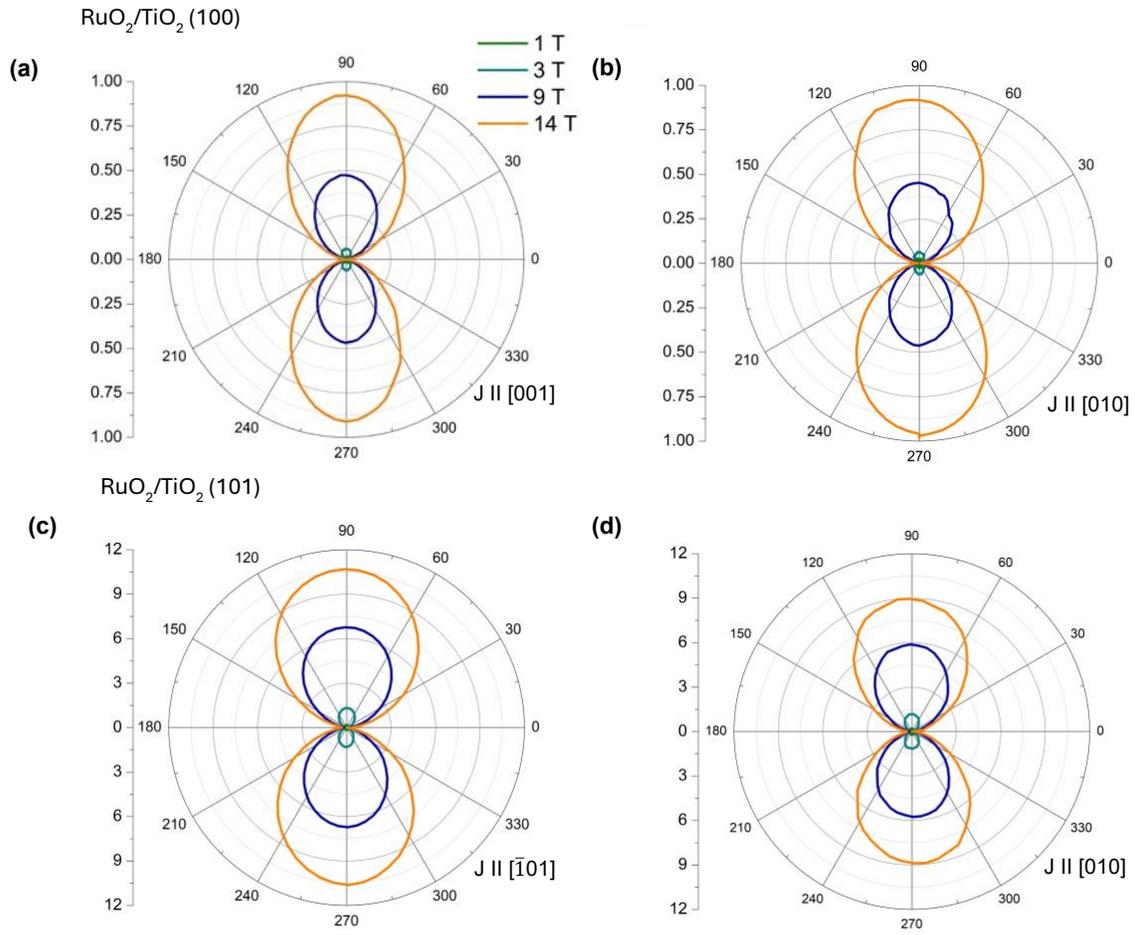

**Figure 4: Anisotropic Magnetoresistance of $RuO_2$ films.** Data collected on $RuO_2$ thin films grown on $TiO_2$ (100) (a,b) and $TiO_2$ (101) (c,d). The results show AMR(%)= $(R_{xx}-R_{min})/R_{min} \times 100$, where the current is kept along a crystallographic direction, using microfabricated Hall bars, while the angle between the current and the applied in-plane magnetic field changes, $\theta = 0° - 360°$. The magnetic field remains unchanged (1 T, 3 T, 9 T, and 14 T) in each revolution. The current direction is rotated 90° degrees between left and right panels using Hall bars along two perpendicular edges of the sample.